 \documentstyle[aps,multicol]{revtex}
\tighten
\draft
\title{Low frequency response
of a collectively pinned vortex manifold}
\author{S. E. Korshunov}
\address{L. D. Landau Institute for Theoretical Physics,
Kosygina 2, Moscow 117940, Russia}
\begin{document}
\maketitle

\begin{abstract}
A low frequency dynamic response of a vortex manifold in a type-II
superconductor can be associated with thermally activated tunneling
of large portions of the manifold between pairs of metastable states.
We suggest that statistical properties of these
states can be verified by using the same approach for the analysis of
thermal fluctuations the behaviour of which is well known.
The exponent describing the frequency dependence of a linear response
is found for the generic case of a vortex manifold with non-dispersive
elastic moduli and also for the case of thin superconducting film
in which the compressibility modulus is always non-local.

$~~$


\end{abstract}

  \begin{multicols}{2}
\section{Introduction}
In various situations the theory of weak collective pinning (for a
review see Ref. \cite{BFGLV}) treats a vortex medium in a type-II
superconductor like an elastic manifold interacting with a random
pinning potential. In particular, a vortex  contribution to a low
frequency response to an applied current (impedance) is ascribed
to the classical (thermally activated) tunneling of large portions
of the vortex manifold between different minima of the
uncorrelated random potential \cite{FFH,FN,KV}.
In the framework of this approach the pairs of states
that allow for thermally activated tunneling between them
are treated as current-biased two-level systems.
Earlier analogous ideas were applied for the description
of the dynamic response of spin glasses \cite{FH} and of randomly
pinned dislocations and interfaces \cite{IV}.

The frequency dependence of the two-level systems contribution to
specific impedance
$z_{tl}(\omega)\equiv\rho_{tl}(\omega)+i{\omega}l_{tl}(\omega)$
of a type-II superconductor penetrated by external magnetic field
has been found
by different groups of authors \cite{FFH,FN,KV} to be of the form:
\begin{eqnarray}
l_{tl}(\omega) & \propto & [\ln(1/\tau_0|\omega|)]^y; \eqnum{1a} \\
\rho_{tl}(\omega) & \propto & |\omega|[\ln(1/\tau_0|\omega|)]^{y-1}.
                                          \label{1} \eqnum{1b}
\end{eqnarray}
but the conclusions of these three groups on the value of the exponent
$y$ are not compatible with each other. \addtocounter{equation}{1}

Comparison shows that the discrepancy appears because the  calculation
of Fisher, Fisher and Huse \cite{FFH}
is based on a semi-phenomenological conjecture that
a contribution  of an  active two-level  system to  inductance can  be
estimated  (from  above)  by  replacing  this two-level system with an
insulating hole  of the  same volume.   This assumption  does not take
into account that the motion of vortices inside of a two-level  system
leads to a  change of a  phase distribution in  superconductor (and of
the energy of electric  current in it) also  outside of the limits  of
this particular two-level system.
Accordingly, it turns out to be in direct
contradiction  with  the  explicit  expression  for a two-level system
contribution to  impedance derived  by Koshelev  and Vinokur \cite{KV}
and used in their calculation of $z_{tl}(\omega)$, and naturally leads
to a different value of $y$.
The less pronounced descrepancy between the results of Ref. \cite{FN}
and Ref. \cite{KV} can be explained, in particular, by the difference
in assumptions on statistics of metastable states.

In the present work we revisit this problem following the more
reliable (as is demonstrated below) approach of Koshelev and Vinokur
\cite{KV}. We start by deriving the value of the exponent
$y$ for the generic case of a vortex manifold with a single
elastic modulus (but with arbitrary dimensionality). Any more complex
system with different (but local) elastic moduli will be
characterized by the same value of $y$. In Ref. \cite{KV}
the value of $y$ in non-dispersive system has been found only
for the particular case of a single vortex pinning.

We then suggest a new independent way to check the validity of the
numerous assumptions involved in calculation of the two-level systems
contribution to a linear dynamic response by using the same set of
assumptions for calculation of another quantity (the amplitude of
thermal  fluctuations), which on the other hand can be calculated
exactly, because in the framework of a random manifold description it
has to be the same as in absence of disorder \cite{SVBO}. It turns
out that application of the assumptions used in the previous
calculation indeed produces an answer which is in agreement with
the well known result for the pure case.

The possibility of the additional check turns out to be very
useful when we address the case of a thin
superconducting film in which the compressibility modulus $c_{11}$
of vortex manifold is always non-local.
It allows to draw some conclusions about
size and shape distribution of two-level systems which otherwise
would be unavailable. Inclusion of these conclusions into
calculation leads [for $c_{11}(q)\propto 1/q^2$] to very weak
(double logarithmic) frequency dependence of $l_{tl}(\omega)$
corresponding to $y=0$.

The case of a thin superconducting film is also of a special
interest in relation with recent experimental investigation of
$z(\omega)$ in ultrathin YBa$_2$Cu$_3$O$_7$ films \cite{Ne,Cal}.
Note that in a two-dimensional (2D) geometry with complete
penetration of the magnetic field the frequency-dependent specific
(sheet) impedance directly determines the response of a film to
external field, whereas in the case of a bulk superconductor it
has to be extracted from the surface impedance.

In all the cases we have considered $l_{tl}(\omega)$ diverges
for $\omega\rightarrow 0$. Thus the important consequence
of our results is that the random manifold approximation can not be
sufficient for the description of a truely superconducting vortex
glass phase \cite{F89,FGLV} with finite superfluid density.

The outline of the article can be summarized as follows. In Sec. II
the random manifold problem is briefly introduced and some
situations when it is appropriate for the description of a vortex
manifold in a superconductor are specified. In Sec. III the
statistical properties of the metastable states which have to be
taken into account in the framework of the two-level system
approach are discussed. In Sec. IV the application of this approach
to calculation of the vortex medium contribution to impedance
is presented in the systematic form for the non-dispersive case.

In Sec. V the same approach is used for the analysis of thermal
fluctuations. We show that (with the same set of assumptions as
used earlier in calculation of a linear dynamic response) it indeed
produces an answer which is consistent with expectations based on
consideration of pure system.

Sec. VI is devoted to the discussion of a thin superconducting film
in which the non-locality of the compressibility modulus is
always important. We show that if one assumes that the dominant
two-level systems in this case are strongly anisotropic (as
suggested by the energy balance estimates used in the analysis
of the non-linear creep \cite{FGLV,FGL,VKK,Kosh}), the expression for
thermal fluctuations amplitude turns out to be convergent in contrast
to its logarithmic divergence in the pure system. The only way to
resolve this contradiction consists in assuming that the
statistics of metastable states is dominated by the presence of
hierarchical sequence of quasi-isotropic two-level systems. The
same distribution is then used for the calculation of the
components of $z_{tl}(\omega)$. In Sec. VII the results are
summarized and discussed.

\section{Random manifold problem}
An elastic manifold (with internal dimension $D$) interacting with
a random pinning potential can be described by the Hamiltonian:
\begin{eqnarray}
H & = & H_{\rm el}+H_{\rm d} \nonumber \\
  & = & \frac{1}{2}\int_{}^{}d^D{\bf x}_1 \int_{}^{}d^D{\bf x}_2
[G^{-1}_0({\bf x}_1-{\bf x}_2)]^{ab}u^a({\bf x}_1) u^b({\bf x}_2)
                                                 \nonumber \\
 & & +\int_{}^{}d^D{\bf x}~v[{\bf x},{\bf u}({\bf x})] \label{a1}
\end{eqnarray}
where the $N$-dimensional vector ${\bf u}({\bf x})\equiv u^a({\bf
x})$ is the displacement of the manifold. The first term in Eq.
(\ref{a1}) describes (in a most general form) the elastic energy
of the manifold and the second the energy of its interaction with
a random pinning potential $v({\bf x},{\bf u})$.

The simplest assumption would consist in assuming that the random
potential $v({\bf x},{\bf u})$ has a Gaussian distribution with
\begin{eqnarray}
\langle v({\bf x},{\bf u})\rangle_{\rm d} & = & 0   \label{a2}\\
\langle v({\bf x}_1,{\bf u}_1)v({\bf x}_2,{\bf u}_2)\rangle_{\rm d} &
= & \delta({\bf x}_1-{\bf x}_2)w({\bf u}_1-{\bf u}_2) \label{a3}
\end{eqnarray}
Here and further on the angular brackets with subscript $d$
stand for the average over disorder,
and with subscript $th$ for the thermal average.
We discuss only the case of a short-ranged random potential
correlation function $w({\bf u})$.

In the simplest situation (which would imply, in particular, full
isotropy and absence of dispersion) the first term in Eq. (\ref{a1})
can be chosen in the form
\begin{equation}
H_{\rm el} =\frac{J}{2}\int d^D{\bf x} \left(\frac{\partial
u^a}{\partial x^\beta}\right)^2                     \label{a4}
\end{equation}
with a single elastic modulus $J$.

The physical systems which can be described by the Hamiltonian of
the form (\ref{a1}) include, in particular, a domain wall in a 2D
or 3D Ising type ferromagnet/antiferromagnet ($D=1,2$; $N=1$); a
dislocation in a crystal ($D=1$, $N=2$); a single vortex line in a
large area Josephson junction ($D=1$, $N=1$) or bulk
superconductor ($D=1$, $N=2$); a vortex medium in superconducting
film ($D=2$, $N=2$) or bulk superconductor ($D=3$, $N=2$), a
layered superconductor with in-plane field ($D=3$, $N=1$) or a
large area Josephson junction with in-plane field ($D=2$, $N=1$).
In all these cases the random pinning potential is automatically
provided by impurities present in any solid or/and by
geometrical inhomogeneities.

Note, however, that the random manifold approximation assumes the
energy of the interaction with the inhomogeneities to be uncorrelated
for different displacements, whereas the energy of the interaction of
an ideal vortex crystal with the inhomogeneities does not change if
the vortex crystal is shifted as a whole by one lattice constant.
Thus, the area of applicability of the random manifold approximation
for the description of vortex crystal pinning is resticted. One can
use this approach if the relevant displacements do not exceed the
lattice period (Larkin regime \cite{L}) or when the ordering
in the vortex crystal is destroyed by the presence of
defects whose motion with respect to the vortex manifold
is dynamically frozen in comparison with
the motion of the manifold itself.

Recent experiments of the Neuch\^{a}tel group on ultrathin
YBa$_2$Cu$_3$O$_7$ films \cite{Cal} have demonstrated a crossover to
the regime in which the contribution to resistivity associated
with the motion of point-like defects is negligible in comparison
with the contribution which can be ascribed to collective pinning
behaviour.

\section{Low energy metastable states}

The low-frequency dynamics of a weakly pinned elastic manifold can
be associated with the thermally activated tunneling of large
domains of the manifold between different minima of the random
potential \cite{IV}. In a simple system with a discrete spectrum
only the tunneling between the ground state and the first excited
state is of importance at low temperatures, thus it can be reduced
to a two-level system. For an infinite manifold one should
take into account that such two-level systems appear at all scales
and form a hierarchical structure, i. e.  if some domain of the
manifold can tunnel between some states one also has to consider
the possibility of tunneling of smaller domains inside this area
between different pairs of states.

Each of such two-level systems can be characterized by its (linear)
size $L$, its volume $V\sim L^D$ (untill specified we discuss the
simplest isotropic case with a single non-dispersive elastic
modulus), the typical vortex displacement between the two
states $u_{}$, which, for example, can be defined by
\begin{equation}
u_{}^2=\frac{1}{V}\int_{}^{}d^D{\bf x}\,{\bf u}^2({\bf x}),
                                                    \label{g8}
\end{equation}
the energy difference between the two states $\Delta$
and the energy barrier $U$ which has to be overcome for moving the
manifold (the vortex bundle) from one of the two states to the other.

The universality hypothesis introduced by Ioffe and Vinokur
\cite{IV} suggests that for each length scale $L$ there should
exist only one relevant energy scale $E(L)$ such
that, in particular, the typical values of $\Delta$ and $U$ for a
system of size $L$ are proportional to $E(L)$. The magnitude of
$E(L)$ can be then estimated by estimating the elastic energy
associated with the displacement of the vortex bundle
of the size $L$:
\begin{equation}
E(L)\sim J V(L)\frac{u^2_{}(L)}{L^2}.
                                                        \label{b1}
\end{equation}
If the scale dependence of the typical displacement $u_{}(L)$
is given by $u(L)\propto L^\zeta$
(where $\zeta$ is usually called the wandering exponent)
Eq. (\ref{b1}) leads to $E(L)\propto L^\chi$ with
\begin{equation}
\chi=2\zeta+D-2.                                        \label{b1a}
\end{equation}
On the other hand Fisher, Fisher and Huse \cite{FFH} have suggested
that the scale dependence of the typical energy barrier $U(L)$
can be described by another exponent $\psi$ not necessariliy
coinsiding with $\chi$ ($\psi\geq\chi$).
For the sake of generality in the following we will keep
(for a while) the separate notation for $\psi$, although we will
assume that the typical value of $\Delta$ (for the given length
scale $L$) can be estimated with the help of Eq. (\ref{b1}).

Various properties of the manifold depend also on the form of the
size distribution function $\nu(L)$ of the two-level systems. The
hierarchical distribution implies that any two-level system can
include smaller two-level systems whose size differ from that of
the "parent" two-level system by some numerical factor of the
order of one. Thus the ratio of the typical "neighboring" length
scales has to be more or less constant across the whole
length-scale range involved \cite{FH}. This is compatible with a
uniform distribution of the logarithms of the length scales.
However, for each length scale one should also include the factor
$1/V(L)$ proportional to the largest possible concentration of
non-overlapping two-level systems of size $L$. Thus $\nu(L)$ has
to be of the form
\begin{equation}
dL~\nu(L)\propto\frac{dL}{L}\frac{1}{V(L)}          \label{b3}
\end{equation}
Koshelev and Vinokur \cite{KV} have introduced the first factor
(imposed by the hierarchical structure) in the r.h.s. of Eq.
(\ref{b3}) in the form $dU/U$ (without any explanation), which for
$U(L)$ algebraically dependent on $L$ is equivalent to $dL/L$, whereas
in Ref. \cite{FN} the hierarchical nature of the size distribution of
two-level systems has not been taken into account.

\section{Two-level systems contribution to impedance}

The contribution of the two-level systems to the specific
impedance of a superconductor is given \cite{KV} by
\begin{equation}
z_{tl}(\omega)=n\frac{\gamma}{T}\left<\frac{(V\overline{u})^2}
{\cosh^2(\Delta/2T)}
\frac{i\omega}{1+i\tau\omega}\right>_{\rm d};~
\gamma=\frac{B^2}{4c^2}                                    \label{c1}
\end{equation}
where $n$ is the concentration of such systems, $B$ is the
magnetic induction and  $T$ is the temperature, whereas $V$,
$\overline{u}$, $\Delta$ and $\tau$
are parameters characterizing a particular two-level system:
$V$ is the volume in which the vortices are displaced
(or the area in the case of a 2D superconductor),
$\overline{u}$ is the average displacement of the vortices inside the
bundle in the direction of current-induced force and  $\Delta$ is the
difference in energy between the two states. The relaxation time
$\tau$ describing the rate of the thermally activated (incoherent)
tunneling between the two states depends exponentially:
\begin{equation}
\tau=\tau_0\exp(U/T)                                     \label{c4}
\end{equation}
on the barrier $U$ separating them.

If one splits $z_{tl}(\omega)$ into real and imaginary parts:
\begin{equation}
z_{tl}(\omega)=\rho_{tl}(\omega)+ i\omega l_{tl}(\omega)  \label{c5}
\end{equation}
and assumes that the average over disorder can be estimated by
taking for each length scale the typical (scale-dependent) values
of all the parameters involved, the expression for the two-level
systems contribution to specific impedance $l_{tl}(\omega)$ is
reduced to
\begin{equation}
l_{tl}(\omega)\sim\frac{\gamma}{T}\int^{L_\omega}_{L_c}dL\,\nu(L)V^2(L)
\overline{u}^2(L)
\langle\cosh^{-2}(\Delta/2T)\rangle_{\rm d} (L)                   \label{c6}
\end{equation}
Due to the exponentially fast increase of $\tau$ with $L$, instead
of including in Eq. (\ref{c6}) the factor
\begin{equation}
\frac{1}{1+[\tau(L)\omega]^2}                           \label{c7}
\end{equation}
the integration in it is cut off (at the upper limit) at a
frequency-dependent length scale $L_\omega$ defined by the relation
$\tau(L_\omega)\omega\sim 1$. For
$U(L)\propto\epsilon(L/L_c)^\psi$
\begin{equation}
L_\omega\propto L_c\left(\frac{T}{\epsilon}
\ln\frac{1}{\tau_0|\omega|}\right)^{1/\psi}           \label{c8}
\end{equation}

The integration interval in Eq.
(\ref{c6}) is limited from below by the (temperature dependent)
collective pinning length $L_c$ \cite{BFGLV} which determines the
boundary between the different regimes of fluctuations.
At length scales lower than $L_c$ the manifold can be considered as
fluctuating within one of the minima of the (thermally renormalized)
random potential, whereas for larger scales only the jumps
between  different valleys of the potential are of importance.
The contribution to $l(\omega)$ from length scales smaller than
$L_c$ has a finite limit for $\omega\rightarrow 0$.

Since $\Delta$ is the difference in energy between the two spatially
separated states in the uncorrelated random potential,
the distribution function $p(\Delta)$
can be expected to  remain finite for $\Delta\rightarrow 0$
(the broad distribution assumption \cite{FH}).
The last factor in the r.h.s. of Eq. (\ref{c6}) (the fraction of
"thermally active" two-level systems) can be then estimated
\cite{FH,IV} as
\begin{equation}
\langle\cosh^{-2}(\Delta/2T)\rangle_{\rm d}(L)
\sim\frac{T}{\Delta_{}(L)} \label{c9}
\end{equation}
where $\Delta_{}(L)$ is the typical value of $\Delta$
for the length scale $L$ [for example the width of $p(\Delta)$].
Note that (for any
scale) only small fraction of two-level systems is assumed to be
not frozen and therefore involved in linear dynamic response (or
thermal fluctuations). Thus they are expected to be well separated
from each other, which justifies neglecting their interaction.

Substitution of Eq. (\ref{c9}) into Eq. (\ref{c6}) leads to
\begin{equation}
l_{tl}(\omega)\sim {\gamma}\int^{L_\omega}_{L_c}dL\,\nu(L)
\frac{V^2(L)\overline{u}^2(L)}{\Delta_{}(L)}    \label{c10}
\end{equation}
According to the universality hypothesis \cite{IV} $\Delta_{}(L)$
has to be of the same order of magnitude as the elastic
contribution to energy estimated in Eq. (\ref{b1}), which for
$\overline{u}\sim u_{}$ gives:
\begin{equation}
\frac{V^2(L)\overline{u}^2(L)}{\Delta_{}(L)}\sim V(L)\frac{L^2}{J}
\label{c11} \end{equation}
Substitution of Eqs. (\ref{b3}) and (\ref{c11}) into Eq.
(\ref{c10}) then leads to
\begin{equation}
l_{tl}(\omega)\propto\frac{\gamma}{J}\int_{L_c}^{L_\omega}\frac{dL}{L}L^2
\propto\frac{\gamma}{J}L_c^2\left(\frac{T}{\epsilon}
\ln\frac{1}{\tau_0|\omega|}\right)^{y}, \label{c12}
\end{equation}
where $y={2}/{\psi}$.

With the same assumptions that have been used for the derivation of
Eq. (\ref{c12}) the two-level system contribution
to the resistivity is given by
\begin{equation}
\rho_{tl}(\omega)\propto\frac{\gamma}{J}\int_{L_c}^{\infty}dL~L
\frac{\tau(L)\omega^2}{1+[\tau(L)\omega]^2}          \label{c13}
\end{equation}
Alternatively $\rho_{tl}(\omega)$ can be restored from
$l_{tl}(\omega)$ with the help of the simplified form \cite{LSB,PI}
of the Kramers-Kronig relation:
\begin{equation}
\rho_{tl}(\omega)\approx -|\omega|\frac{\pi}{2}
\frac{d}{d\ln|\omega|} l_{tl}(\omega)                \label{c14}
\end{equation}
which is applicable for $l_{tl}(\omega)=f(\ln|\omega|)$.
Both methods give
\begin{equation}
\rho_{tl}(\omega)\propto\frac{\gamma}{J}
\left(\frac{T}{\epsilon}\right)^{y}|\omega|
\left(\ln\frac{1}{\tau_0|\omega|}\right)^{y-1} \label{c15}
\end{equation}

It can be shown that $y$ is equal to $2/\psi$ not only for
the simplest case of a single elastic modulus, but for the
general case of non-dispersive moduli. In a bulk superconductor at
large enough scales (which corresponds to low enough frequencies)
all elastic moduli become local. The case of a thin film in which
the strong dispersion of the compressibility modulus is
unavoidable is considered in Sec. VI.

For finite values of the current density $j$ Eq. (\ref{c1}),
which has been the starting point of our calculation, is
applicable only if $V\overline{u}B/c$ is small in comparison with
temperature.
This determines the current dependence of the length scale
$L_j\propto (T/j)^{\frac{1}{D+\zeta}}$ at which the integration
in Eq. (\ref{c6}) should be cut off if $L_j\ll L_\omega$.
In that case the growth of $l_{tl}(\omega)$
with decreasing $\omega$ saturates at
$l_{tl}(\omega=0,j)\propto (T/j)^\frac{2}{D+\zeta}$.

\section{Comparison of two approaches to calculation
of thermal fluctuations amplitude}

In the present work we suggest an independent way to check the
validity of the two-level system approach for the description of
the linear dynamic response of a collectively pinned manifold. This
can be done because in the random manifold problem the static
irreducible correlation function
\begin{eqnarray}
\langle\langle u^a({\bf x}_1) u^b({\bf x}_2)\rangle\rangle &
\equiv & \langle\langle [u^a({\bf x}_1)-\langle u^a({\bf
x}_1)\rangle_{\rm th}]                \nonumber \\
 & & \times[u^b({\bf x}_2)-\langle u^b({\bf
x}_2)\rangle_{\rm th}] \rangle_{\rm th}\rangle_{\rm d}      \label{d1}
\end{eqnarray}
(which can be associated with thermal fluctuations) according to
Shultz {\em et al} \cite{SVBO} should be exactly the same as in
the absence of disorder:
\begin{equation}
\langle\langle u^a({\bf x}_1) u^b({\bf x}_2)\rangle\rangle
=T G^{ab}_0({\bf x}_1-{\bf x}_2).                 \label{d2}
\end{equation}
(a brief derivation can be found in Appendix). An analogous
relation for the case of periodic behaviour of $w({\bf u})$ with
respect to displacement has been suggested by Dotsenko and
Feigel'man \cite{DF}.

In the presence of disorder the long-distance behaviour of the
correlation function (\ref{d1}) has to be mediated by the
two-level systems. Therefore, the investigation of thermal
fluctuations in terms of the two-level system approach and
comparison of the result with well known result for the pure
system allows to check the validity of different conjectures
involved in the calculation of a linear dynamic response. Instead
of considering the dependence of $\langle\langle u^a({\bf
x}_1)u^b({\bf x}_2)\rangle\rangle$ on $|{\bf x}_1-{\bf x}_2|$ one
can alternatively investigate the dependence of $\langle\langle
u^2\rangle\rangle$ on the size of the system $L_{0}$, which also
has to be the same as in the pure case.

For a single two-level system with the energy gap $\Delta$
the amplitude of the thermal fluctuations of the displacement
is given by
\begin{equation}
\langle\langle u^2\rangle\rangle
\equiv\langle(u-\langle u\rangle_{\rm th})^2\rangle_{\rm th}
=\frac{(u_1-u_2)^2}{4\cosh^2(\Delta/2T)}     \label{e1}
\end{equation}
For a collectively pinned manifold the dominant large-scale
contribution to $\langle\langle u^2({\bf x)}\rangle\rangle$ should
come from the two-level systems which include the point ${\bf
x}$ and [on the same assumptions as have been used while calculating
$l_{tl}(\omega)$ and $\rho_{tl}(\omega)$] can be estimated as
\begin{equation}
\langle\langle u^2_{tl}\rangle\rangle\sim T \int_{L_c}^{L_{0}}
dL\,\nu(L) \frac{V(L) u^2_{}(L)}{\Delta_{}(L)} \label{e2}
\end{equation}
where the upper limit of integration is now imposed by the size of
the system and [in accordance with Eq. (\ref{c9})]
$\langle\cosh^{-2}(\Delta/2T)\rangle_{\rm d}$ has been already
replaced with $T/\Delta_{}(L)$.

Substitution of Eqs. (\ref{b1}) and  (\ref{b3}) into
Eq. (\ref{e2}) then gives
\begin{equation}
\langle\langle u^2_{tl}\rangle\rangle\propto
\frac{T}{J}\int_{L_c}^{L_{0}}\frac{dL}{L^{D-1}}\label{e3}
\end{equation}
which, if compared with the  trivial result for the case without
disorder (for which $\langle\langle u^2\rangle\rangle\equiv\langle
u^2\rangle_{\rm th}$):
\begin{equation}
\langle\langle u^2\rangle\rangle\approx
\frac{T}{J}\int_{|q^\alpha|>{\pi}/{L_{0}}}
\frac{d^D{\bf q}}{(2\pi)^D}\frac{1}{q^2}      \label{e5}
\end{equation}
reproduces all its important features.
Namely, the r.h.s. of Eq. (\ref{e3})
(i) contains the correct prefactor $T/J$;
(ii) demonstrates the correct dependence on the size of the system:
\begin{equation}
\langle\langle u^2\rangle\rangle\propto\left\{\begin{array}{lr}
L^{2-D}_{0} & D<2 \\
\ln(L_{0}) & D=2 \\
L_c^{2-D}- L^{2-D}_{0} & D>2 \end{array}\right.                                 \label{e4}
\end{equation}
and (iii) the system-size dependent contribution to
$\langle\langle u^2_{tl}\rangle\rangle$
does not depend on the unknown disorder-related parameters $L_c$
and $\epsilon$.

This allows to conclude that different assumptions which have been
used while calculating $l_{tl}(\omega)$, $\rho_{tl}(\omega)$
and $\langle\langle u^2_{tl}\rangle\rangle$
[the universality hypothesis, the hierarchical distribution of
two-level systems, the broad distribution assumption for $p(\Delta)$]
were indeed chosen in a reasonable way.

\section{Thin superconducting film in perpendicular magnetic field}

The long-range interaction of vortices makes the compressibility
modulus $c_{11}$ of a bulk superconductor strongly non-local for
the wave-lengths smaller than the magnetic field penetration depth
$\lambda$. In a thin superconducting film the penetration depth
$\Lambda$ is strongly increased in comparison with that of a bulk
superconductor: $\Lambda=2\lambda^2/d$ \cite{P}, where $d\ll
\lambda$ is the thickness of the film. Therefore, in a thin film
the dependence $c_{11}({q})\approx \overline{c}_{11}/q^2$ (where
$\overline{c}_{11}\approx{B^2}/{2\pi\Lambda}\propto{c_{66}}a^{-2}$)
resulting from a non-screened vortex-vortex interaction holds in a
much wider range of length scales than in a bulk superconductor.
Here $c_{66}\approx\Phi_0 B/32\pi^2\Lambda$ is the shear modulus
of the film (which in contrast to $c_{11}$ is always local),
$\Phi_0=hc/2e$ is the flux quantum and $a^{-2}=B/\Phi_0$ is the
vortex density.

If one tries to shift a vortex bundle in such a system
(in search of the next potential minimum),
it turns out that the optimal shape of the bundle
is strongly anisotropic \cite{FGLV,FGL,VKK,Kosh}
with the size in the direction of the displacement $L$ much larger
than the size in the perpendicular direction $L_\perp$.
The optimal relation between $L$ and $L_\perp$ can be found
by minimizing the total elastic energy for the given area
of a bundle $S\propto L L_\perp$.
Minimization (for fixed $S$) of $E_{\rm com}+E_{\rm sh}$ where \cite{Kosh,VKKc}
\begin{equation}
E_{\rm com}\sim
\overline{c}_{11}{S^2}\left(\frac{\overline{u}}{L}\right)^2 \label{g4}
\end{equation}
and
\begin{equation}
E_{\rm sh}\sim  c_{66}S\left(\frac{u_{}}{L_\perp}\right)^2 \label{g3}
\end{equation}
or a simple comparison of $E_{\rm com}$ with $E_{\rm sh}$ for
$\overline{u}\sim u_{}$ gives
\begin{equation}
L_{}\propto \frac{L_\perp^3}{a^2}\gg L_{\perp}.      \label{g5}
\end{equation}
Note, however, that when the energy scale defined by Eq. (\ref{g3})
is used as an estimate for $\Delta_{}$,
the factor $Su^2_{}/\Delta_{}$ in the 2D version of Eq. (\ref{e2})
is reduced to
\begin{equation}
\frac{Su^2_{}}{\Delta_{}}\sim\frac{L_\perp^2}{c_{66}}
                                                     \label{g5a}
\end{equation}
for arbitrary relation between $L$ and $L_{\perp}$.

The calculation of Sec. V has confirmed that the form of the size
distribution of two-level systems can be correctly estimated by
assuming that they do not overlap with each other, but can
be situated inside each other forming hierarchical structures.
Such estimate is consistent with the conjecture that the number of
metastable states to which a particular domain of a manifold can
tunnel (without moving a much larger part of the manifold) is always
of the order of one \cite{IV}.
In what follows we assume that the same property holds also in
presence of dispersion.

The most optimistic estimate for $\nu(L)$ can be then obtained by
assuming that for all scales the strongly anisotropic two-level
systems are arranged in the most advantageous way to cover all the
area available, which corresponds to
\begin{equation}
dL\,\nu(L) \Rightarrow \frac{dL}{L}
\frac{1}{L L_\perp}.                                 \label{g6}
\end{equation}
The more realistic estimate should probably take into account
that independent anisotropic two-level systems are likely to have
uncorrelated orientations, so  the requirement of non-overlapping
will lead to $\nu(L)\propto L^{-3}$.

After substitution of Eq. (\ref{g5a}) into the integral [of the form
(\ref{e2})] defining $\langle\langle u^2_{tl}\rangle\rangle$
one obtains that for $L\propto L_\perp^3$ it is convergent at
the upper limit:
\begin{equation}
\langle\langle u^2_{tl}\rangle\rangle\propto\frac{T}{c_{66}}
\int_{L_c}^{\infty} \frac{dL}{L}\left(\frac{L_\perp}{L}\right)
<\infty                                              \label{g7}
\end{equation}
even for the optimistic form of $\nu(L)$ given by Eq. (\ref{g6}).
On the other hand, we know that in a pure system the contribution
of the transverse modes (which depends only on the shear modulus
which is local) leads to the logarithmic divergence of
$\langle\langle u^2\rangle\rangle\equiv\langle u^2\rangle_{\rm th}$.
It follows from the results of Schulz {\em et al} \cite{SVBO} that
in the presence of disorder the same behaviour should be mediated
by the large-scale two-level systems.

A plausible way to explain the logarithmic divergence of
$\langle\langle u^2_{tl}\rangle\rangle$ consists in assuming that
the film should contain not only anisotropic two-level systems with
$L\gg L_\perp$, but also an hierarchical sequence of quasi-isotropic
two-level systems in which $L_\perp$ is of the same order as $L$.
For such two-level systems the requirement of the balance between
the different contributions to the elastic energy
($E_{\rm com}\sim E_{\rm sh}$) leads to $\overline{u}\propto(a/L)u_{}\ll
u_{}$, which means that the displacement of the
vortices in quasi-isotropic bundles is mostly of rotational type.

The linear dynamic response has to be associated with the same
degrees of freedom as are taken into account in the calculation of
the static thermal fluctuations. However, $z_{tl}(\omega)$ can not
be obtained by direct application of the fluctuation-dissipation
theorem to $\langle\langle u^2_{tl}\rangle\rangle$, since these
two quantities include different linear combinations of the
degrees of freedom involved. Nonetheless, when calculating
$z_{tl}(\omega)$ one should take into account the same set of
two-level systems as for the calculation of $\langle\langle
u^2_{tl}\rangle\rangle$, in contrast to the case of the non-linear
creep for which the shape of the moving vortex bundles is imposed
by the applied current \cite{FGL,VKK,Kosh}.

Application of the expression (\ref{g4}) for $E_{\rm com}$ as an
estimate for $\Delta_{}$ shows that the factor
$S^2 \overline{u}^2/\Delta_{}$ in the 2D version of Eq. (\ref{c10})
does not depend on $L_\perp$ and can be estimanted as
$L^2/\overline{c}_{11}\sim L^2\Lambda/B^2$.
For the hierarchical sequence of quasi-isotropic  two-level
systems with $\nu(L)\propto L^{-3}$ this leads to the extremely weak
(double logarithmic) frequency dependence of
\begin{equation}
l_{tl}(\omega)\propto\frac{\gamma\Lambda}{B^2}\int_{L_c}^{L_\omega}
\frac{dL}{L}
\propto \frac{\Lambda}{c^2}\ln\left(\frac{T}{\epsilon}
\ln\frac{1}{\tau_0|\omega|}\right),                    \label{g11}
\end{equation}
which can hardly be expected to be resolvable from the background
superfluid contribution $l_0$ in the experiments probing the
low-frequency response of thin films.

However, substitution of Eq. (\ref{g11}) into Eq. (\ref{c14})
gives
\begin{equation}
\rho_{tl}(\omega)\propto \frac{\Lambda}{c^2}
\frac{|\omega|}{\ln(1/\tau_0|\omega|)}                \label{g12}
\end{equation}
which, in contrast to Eq. (\ref{g11}), does not exhibit
any specially weak dependence on $\omega$.
Note that two unknown disorder-related parameters $L_c$
and $\epsilon$ as well as the magnetic field dependence have
dropped out from Eq. (\ref{g12}).

The presence of the hierarchical sequence of quasi-isotropic
two-level systems still leaves enough place for more optimal
anisotropic two-level systems with $L_\perp\ll L$. Although they
do not contribute much to $\langle\langle
u^2_{tl}\rangle\rangle$, their contributions to the components of
$z_{tl}(\omega)$ could be of importance. However, their size
distribution will be forced by the presence of hierarchical
sequence of quasi-isotropic two-level systems to be of the same
form $\nu(L)\propto L^{-3}$ and, therefore, their contribution to
$l_{tl}(\omega)$ and $\rho_{tl}(\omega)$ will be of the same form
as given by Eqs. (\ref{g11})-(\ref{g12}).

In a thin superconducting film the compressibility modulus
$c_{11}$ is non-local not only for $\Lambda q\gg 1$ where
$c_{11}({ q})\approx \overline{c}_{11}/q^2$, but also  for
$\Lambda{ q}\ll 1$ where $c_{11}({ q})\approx B^2/2\pi q$. An
analogous calculation for such form of $c_{11}({ q})$ produces for
the components of $z_{tl}(\omega)$ the answers of the form
(1) with $y=1/\psi$.

\section{Conclusion}

In the present work we argue that thermal fluctuations of a
collectively pinned vortex manifold are determined by the same
degrees of freedom (related to thermally activated tunneling
between the pairs of low-lying metastable states - two-level
systems) as its low-frequency linear dynamic response. Therefore
one can use the known dependence of thermal fluctuations amplitude
on the size of the system (which has to be exactly the same as
in absence of disorder \cite{SVBO}) for checking the consistency of
the conjectures which are used in the calculation of the linear
dynamic response. The set of the assumptions which are necessary to
produce the correct answer for the thermal fluctuations amplitude
includes, in particular, the conjecture on hierarchical
distribution of two-level systems (which means that they can be
situated inside each other) and also the universality hypothesis.
If the same set of assumptions is used for the calculation of a
vortex manifold contribution to impedance its frequency dependence
(in absence of dispersion) is given by Eqs. (\ref{1}) with
$y=2/\psi$, where $\psi$ is the exponent (which depends both on $D$
and $N$) describing the scale dependence of the typical energy
barrier $U(L)$.

The same result is also applicable in the limit af small fields,
when one can neglect the interaction between different vortices
and treat each vortex separately as 1D manifold.
In that case one should take the value of $\psi$ corresponding
to $D=1$.

Note that in the framework of our analysis the universality
hypothesis has been used only for the estimate of $\Delta(L)$.
If (as suggested by Ioffe and Vinokur \cite{IV})
it is further assumed that the same energy scale can be
used for the estimate of $U(L)$, the value of $\psi$ will coinside
with $\chi$ given by Eq. (\ref{b1a}).
Different approaches including scaling arguments \cite{FGLV,Kar},
functional renormalization group
\cite{F,H-H} and a self-consistent calculation incorporating replica
symmetry breaking \cite{MP} lead to
\begin{equation}
\zeta=\frac{4-D}{4+\beta N}                             \label{b2}
\end{equation}
with $1/2\leq\beta\leq 1$.
For the case of thin superconducting film (D=2, N=2)
or bulk superconductor (D=3, N=2) Eqs. (\ref{b1a}) and (\ref{b2})
give the values of $\chi$ in the interval from $2/3$ to $7/5$,
that is around $1$.

Koshelev and Vinokur \cite{KV} have found the value
of the exponent $y$ only for the particular values of
$\zeta$ and $\chi$ corresponding to $D=1$ and $N=2$
and not in the general form as above.
In Ref. \cite{FN} the hierarchical nature of the size distribution
(of the two-level systems) has not been taken into account
and the estimate of $\rho_{tl}(\omega)$ has been obtained without
integrating over the scales, which has led (in our notation)
to $y=2/\psi+1$.
As has been already mentioned in the
Introduction the analogous calculation in Ref. \cite{FFH} has
been performed using the assumption which is in contradiction with Eq.
(\ref{c1}) and therefore can not be used for comparison.

Note that $l_{tl}(\omega)$ diverges in the limit of
$\omega\rightarrow 0$, which corresponds to supression of superfluid
density [inversely proportional to $\lim_{\omega\rightarrow 0}l(\omega)$].
Thus the results of this work are not in agreement with the popular
point of view that random manifold approach provides an exhaustive
description of dynamic properties of a truely superconducting
vortex glass phase (which is supposed to be formed at low temperatures
due to pinning \cite{F89,FGLV}),
at least if superconductivity is understood as the ability to carry
a superconducting (non-dissipative) current and not only as the
vanishing of the linear resistance.
Our analysis suggests that in the framework of random manifold approach
the finite value of $l_{tl}(\omega\rightarrow 0)$
is incompatible with the correct scale dependence
of $\langle\langle u^2_{tl}\rangle\rangle$.
Therefore one has to conclude that the accurate description of a
vortex glass phase which can carry a superconducting current
(if such phase exists at all) requires
a more sophisticated treatment than the description of vortex medium
in terms of an elastic manifold interacting with a random potential.
For example, it possibly should take into account that some of the
defects of a vortex lattice are generated by a disorder and can
not freely move with a vortex manifold.

However, both in the case of Larkin regime and in the case of
dynamically frozen thermally excited defects
(frozen vortex liquid regime) the frequency range
of the applicability of the random manifold description of a
vortex medium  in a superconductor is anyway restricted from below.
Moreover, the two-level system contribution to impedance
$l_{tl}(\omega)$ is only logarithmic in
$\omega$ and in a practical situation may be negligible in
comparison with "bare" impedance $l_0$ down to exponentially low
frequencies.

On the other hand $\rho_{tl}(\omega)$ produces in the low
frequency limit the dominant contribution to the resistivity in
comparison with the contributions related with the normal channel
conductance and with the oscillations of manifold within each
minimum of a random potential, both of which at low frequencies
are proportional to $\omega^2$.

In thin superconducting films the compressibility modulus of vortex
manifold is always non-local. This leads to the strong anisotropy of
the vortex bundles participating in the non-linear creep
\cite{FGLV,FGL,VKK,Kosh}.
However our analysis has shown that the correct length-scale
dependence of thermal fluctuations amplitude requires the presence
of a hierarchical sequence of quasi-isotropic two-level systems. A
linear dynamic response (which has to be calculated assuming that
the applied current does not change the properties of the system)
produced by the same set of the two-level systems corresponds to
$y=1/\psi$ for $c_{11}(q)\propto 1/q~(\Lambda q\ll 1)$ and to
$y=0$ [with $l_{tl}(\omega)$ still diverging at
$\omega\rightarrow 0$ but only as a double logarithm] for
$c_{11}(q)\propto 1/q^{2}~(\Lambda q\gg 1)$. The contribution from
the more optimal anisotropic two-level systems can be expected to
be of the same form. In both regimes [$c_{11}(q)\propto 1/q$ and
$c_{11}(q)\propto 1/q^2$] the value of the magnetic field $B$
drops out from the expression for $z_{tl}(\omega)$.

Although in thin films the dc resistivity is always finite due to
thermally activated motion of point-like defects of vortex lattice
(vacancies, interstitials, dislocation pairs) \cite{FGL,VKK}, the
collective pinning behaviour has been found to be accessible to
experimental observation \cite{Cal} in the range of
frequencies/temperatures where the activated contribution to
resistivity is too small.

\acknowledgements

This work has been supported in part by the Program "Scientific
Schools of the Russian Federation" (grant No. 00-15-96747) and by
the Swiss National Science Foundation. The author is grateful to
G. Blatter, V. B. Geshkenbein and P. Martinoli for interesting
discussions and useful comments.

\appendix
\section*{}
In the case of a random manifold described by Eqs.
(\ref{a1})-(\ref{a3}) it is possible to show that the irreducible
correlation function of the form (\ref{d1}) remains the same as in
absence of disorder \cite{SVBO}. To this end one can express this
correlation function through the second derivative
\begin{equation}
\langle\langle u^a({\bf x}_1) u^b({\bf x}_2)\rangle\rangle
=T\left.\frac{\partial^2 \tilde{F}} {\partial{s}^a_1
\partial{s}^b_2}\right| _{s^a_1=s^a_2=0}         \label{d3}
\end{equation}
of the (disorder-averaged) free energy
\begin{equation}
\tilde{F}=-T\left\langle\ln\left[\int_{}^{}d{\bf u}
\exp\left(-\frac{\tilde{H}}{T}\right)\right]\right\rangle_{\rm d}
\label{d4}
\end{equation}
with respect to the coefficients in the auxilary (source) terms
added to the Hamiltonian
\begin{equation}
\tilde{H}=H+s^a_1 u^a({\bf x}_1) -s^a_2 u^a({\bf x}_2). \label{d5}
\end{equation}

In order to calculate $\tilde{F}$ it is convenient to shift the
variables ${\bf u}({\bf x})$ (over which the integration in the
partition function is performed) according to
\begin{equation}
u^a({\bf x)}\Rightarrow u^a_*({\bf x})\equiv u^a({\bf x})
+G_0^{ab}({\bf x}-{\bf x}_1)s^b_1 -G_0^{ab}({\bf x}-{\bf
x}_2)s^b_2,       \label{d6}
\end{equation}
which allows to split the non-random contribution to
$\tilde{H}\{u^a\}$ into two terms:
$H_{\rm el}\{u^a_*\}+E(s^a_1,s^a_2)$, the first of which does not
depend on $s^a_{1,2}$ and the second
\begin{eqnarray}
E(s^a_1,s^a_2) & = & -\frac{1}{2}G_0^{ab}(0)s^a_1s^b_1 +G_0^{ab}({\bf
x}_1-{\bf x}_2)s^a_1s^b_2                    \nonumber \\
 & & -\frac{1}{2}G_0^{ab}(0)s^a_2s^b_2
\end{eqnarray}
does not depend on $u_*^a({\bf x})$.

On the other hand the distribution function of the random
potential $v({\bf x}, {\bf u})$ is not affected by the shift
defined by Eq. (\ref{d6}). Therefore the free energy defined by
Eq. (\ref{d4}) differs from its value for original problem (that
is for $s^a_1=s^a_2=0$) only by addition of the term
$E(s^a_1,s^a_2)$, differentiation of which leads to Eq.
(\ref{d2}).


 \end{multicols}

\begin{thebibliography}{99}
\bibitem{VKKc} The analogous estimate for $E_{\rm com}$ in Ref. \cite{Kosh}
              differs from Eq. (\ref{g4}) by the additional
              logarithmic factor, which we believe to be unjustified.
              If the size of a moving bundle is small in comparison
              with the screening length of the logarithmic vortex-vortex
              interaction, the compression energy associated with
              the displacement can not depend on this length.
              This can be shown, for example, by integration by parts
              in Eq. (4) of Ref. \cite{Kosh}.
\end{thebibliography}
\end{document}